\documentclass[twocolumn,pra,superscriptaddress,amsmath,amssymb]{revtex4-1}
\usepackage{graphicx,bm}
\usepackage[colorlinks=true,citecolor=blue,urlcolor=blue]{hyperref}
\usepackage{soul}

\begin{document}
\title{Topology-Enhanced Nonreciprocal Scattering and Photon Absorption in a Waveguide}
\author{Wei Nie}
\affiliation{Institute of Microelectronics, Tsinghua University, Beijing 100084, China}
\affiliation{Frontier Science Center for Quantum Information, Beijing, China}
\affiliation{Theoretical Quantum Physics Laboratory, RIKEN Cluster for Pioneering Research, Wako-shi, Saitama 351-0198, Japan}
\author{Tao Shi}\email{tshi@itp.ac.cn}
\affiliation{Institute of Theoretical Physics, Chinese Academy of Sciences, P.O.
Box 2735, Beijing 100190, China}
\affiliation{CAS Center for Excellence in Topological Quantum Computation,
University of Chinese Academy of Sciences, Beijing 100049, China}
\author{Franco Nori}
\affiliation{Theoretical Quantum Physics Laboratory, RIKEN Cluster for Pioneering Research, Wako-shi, Saitama 351-0198, Japan}
\affiliation{Physics Department, The University of Michigan, Ann Arbor, Michigan 48109-1040, USA}
\author{Yu-xi Liu}\email{yuxiliu@mail.tsinghua.edu.cn}
\affiliation{Institute of Microelectronics, Tsinghua University, Beijing 100084, China}
\affiliation{Frontier Science Center for Quantum Information, Beijing, China}

\begin{abstract}
Topological matter and topological optics have been studied in many systems, with promising applications in materials science and photonics technology. These advances motivate the study of the interaction between topological matter and light, as well as topological protection in light-matter interactions. In this work, we study a waveguide-interfaced topological atom array. The light-matter interaction is nontrivially modified by topology, yielding novel optical phenomena. We find topology-enhanced photon absorption from the waveguide for large Purcell factor, i.e., $\Gamma/\Gamma_0\gg 1$, where $\Gamma$ and $\Gamma_0$ are the atomic decays to waveguide and environment, respectively. To understand this unconventional photon absorption, we propose a multi-channel scattering approach and study the interaction spectra for edge- and bulk-state channels. We find that, by breaking inversion and time-reversal symmetries, optical anisotropy is enabled for reflection process, but the transmission is isotropic. Through a perturbation analysis of the edge-state channel, we show that the anisotropy in the reflection process originates from the waveguide-mediated non-Hermitian interaction. However, the inversion symmetry in the non-Hermitian interaction makes the transmission isotropic. At a topology-protected atomic spacing, the subradiant edge state exhibits huge anisotropy. Due to the interplay between edge- and bulk-state channels, a large topological bandgap enhances nonreciprocal reflection of photons in the waveguide for weakly broken time-reversal symmetry, i.e., $\Gamma_0/\Gamma\ll 1$, producing complete photon absorption. We show that our proposal can be implemented in superconducting quantum circuits. The topology-enhanced photon absorption is useful for quantum detection. This work shows the potential to manipulate light with topological quantum matter.
\end{abstract}

\maketitle

\section{Introduction}
Symmetry-protected topological phases of matter is a growing a field in materials science~\cite{RevModPhys.82.3045,RevModPhys.83.1057,ando2013topological,goldman2016topological,zhang2019topological}, and might find applications in quantum computation~\cite{pachos2012introduction} and quantum technologies~\cite{dennis2002topological,jiang2008anyonic,PhysRevX.10.031041}. In 2008, Refs.~\cite{PhysRevLett.100.013904,PhysRevA.78.033834} proposed to manipulate photon transport using topology, which paved the way for topological photonics~\cite{lu2014topological,Bliokh2015,bliokh2015spin,bliokh2019topological,RevModPhys.91.015006}. In two- or higher-dimensional topological materials, photons can be guided via channels supported by edge states and surface states~\cite{PhysRevLett.100.013905,lu2013weyl,he2016photonic,shi2017topological,PhysRevLett.121.023901}. Due to the large bandgap separating chiral edge states and bulk states, such transport is immune to imperfections, randomness and disorder, and has been realized in different incarnations of optical systems~\cite{fang2012realizing,khanikaev2013photonic,
rechtsman2013photonic,hafezi2013imaging,zilberberg2018photonic,yang2019realization}. The topological protection of photon transport usually takes advantage of the edge states. However, the role of bulk states has not been sufficiently explored.

Light-matter interaction is a fundamental mechanism in quantum physics~\cite{haroche2006exploring}. One-dimensional (1D) waveguides are essential light-matter interfaces and have fundamental applications in quantum devices and quantum  networks~\cite{lodahl2017chiral,RevModPhys.89.021001,RevModPhys.90.031002}. The photon transport in a waveguide can be controlled by coupling to a single atom~\cite{PhysRevLett.95.213001,shen2005coherent,chang2007single,PhysRevLett.101.100501,PhysRevA.78.063827,PhysRevA.81.042304,witthaut2010photon,PhysRevA.82.063816,
PhysRevA.89.053813,PhysRevA.96.053832,PhysRevLett.120.153602,PhysRevA.97.062318,
PhysRevLett.122.073601} or an atom array~\cite{PhysRevA.78.063832,PhysRevA.88.043806,PhysRevA.89.031804,PhysRevA.90.063816,
PhysRevLett.115.163603,PhysRevA.96.043872,PhysRevLett.121.143601}. In the subwavelength regime, the interference of photons emitted from atoms at different positions~\cite{caneva2015quantum,PhysRevA.92.053834,PhysRevLett.117.143602,PhysRevLett.120.140404} gives rise to the collective enhancement of photon transport~\cite{PhysRevA.92.023806,PhysRevA.94.043844,PhysRevLett.123.253601} and directional photon emission~\cite{mitsch2014quantum,PhysRevLett.124.093601,gheeraert2020bidirectional}. In waveguide quantum electrodynamics (QED) systems, e.g., atoms trapped around nanofibers~\cite{okaba2014lamb,PhysRevLett.115.063601,solano2017super}, the direct atom-atom interaction is in general negligible. However, the direct interaction between atoms is essential in superconducting quantum circuits. By engineering this interaction, one can simulate many models in condensed matter physics and high energy physics, including spin models~\cite{johnson2011quantum,barends2016digitized,wang2019synthesis}, lattice gauge theories~\cite{marcos2014two,PhysRevLett.115.240502}, and  topological matter~\cite{PhysRevLett.123.080501,PhysRevLett.124.023603,besedin2020topological}.

\begin{figure*}[t]
\includegraphics[width=18cm]{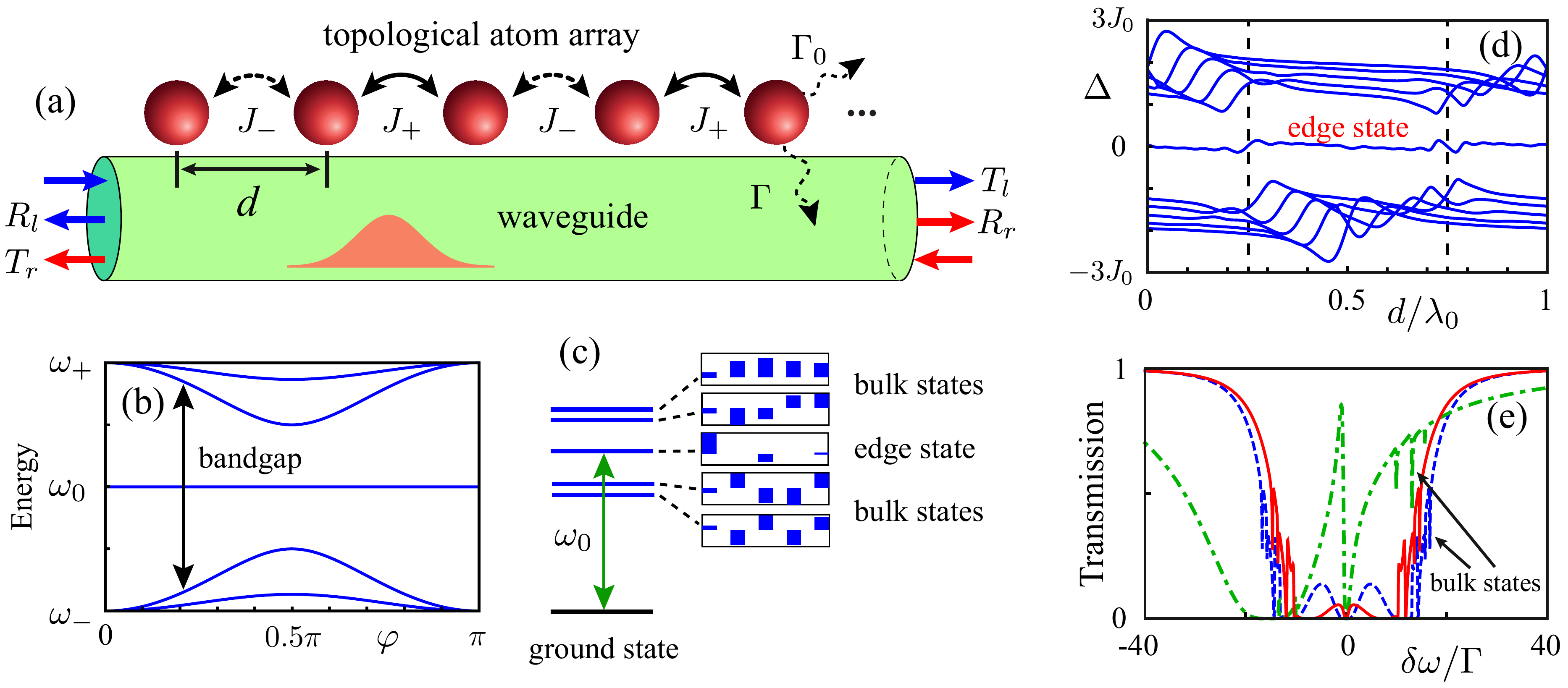}
\caption{(a) Schematic of a 1D waveguide coupled to a topological atom array. The dimerized interactions are $J_{\mp}=J_0(1\mp \cos\varphi)$. Here we consider an array with an odd number of atoms and $0\leq\varphi<\pi/2$. Thus, an edge state is localized at the left edge of the atom array. A homogenous spacing $d$ is assumed for neighboring atoms along the waveguide. All the atoms have decays to the waveguide and environment, denoted by $\Gamma$ and $\Gamma_0$. Here, $T_l$ $(T_r)$ and $R_l$ ($R_r$) represent the transmission and reflection for the left- (right-) incident photon, respectively. (b) Energy spectrum of the SSH array with $5$ atoms. Here, $\omega_{\pm}=\omega_0 \pm 2J_0$. (c) Wave functions of eigenmodes of the SSH atom array in (b) with $\varphi=0.3\pi$. (d) Real energy spectrum $\Delta=\mathrm{Re}(E)$ of the effective Hamiltonian $H_{\mathrm{eff}}$ in Eq.~(\ref{effH}). The edge state is protected from bulk states by the bandgap. The two vertical dashed lines indicate atomic spacings $d=\lambda_0/4$ and $d=3\lambda_0/4$, respectively. (e) Transmission spectra $T=|t(\delta\omega)|^2$ for $d=\lambda_0/4$ (blue-dashed), $d=\lambda_0/2$ (green-dot-dashed) and $d=3\lambda_0/4$ (red-solid). Dips around $\delta\omega=\pm2J_0$ indicate bulk states. In (d) and (e), we consider $J_0/\Gamma=8, \varphi=0.3\pi, \Gamma_0/\Gamma=0.05$, and atom number $N=11$. }\label{fig1}
\end{figure*}

In this Letter, by virtue of the interaction between light and a topological atom array via a waveguide, we show topology-enhanced nonreciprocal scattering. Arrays with an odd number of equally spaced atoms in the Su-Schrieffer-Heeger (SSH) model~\cite{PhysRevLett.42.1698} display broken inversion symmetry. The topological bandgap and edge state yield anomalous photon transport, i.e., complete absorption of light from the waveguide. Distinctive to nontopological atom arrays, the topological atom array enables complete photon absorption for $\Gamma_0/\Gamma\ll1$, where $\Gamma$ and $\Gamma_0$ denote decays to waveguide and parasitic modes in the environment, respectively, as shown in Fig.~\ref{fig1}(a). We employ the so-called ``multi-channel scattering" approach to study the light-matter interaction in the scattering process. By means of this method, we can pinpoint roles of distinctive many-body states in the optical response. We find that for a specific atomic spacing $d$, the edge-state channel is protected and has huge reflection anisotropy. The nonreciprocal reflection is attributed to the interplay between the inversion symmetry breaking induced by the non-trivial topology and the time-reversal symmetry breaking due to dissipation. More precisely, the destructive quantum interference of electromagnetic waves reflected by the dissipative edge- and bulk-state channels gives rise to the \emph{large nonreciprocity}. The topology-protected optical nonreciprocity is beneficial for photon detection with long coherence atoms.

\section{Model}
The photon scattering by independent atoms can be solved by, e.g., transfer matrix~\cite{PhysRevA.78.063832,PhysRevA.90.063816} and input-output methods~\cite{PhysRevA.88.043806,caneva2015quantum}. For the atom array with strong internal interactions studied here, collective modes with unique spectrum structure are pivotal to understand the photon scattering. Hence, we present a multi-channel scattering method in terms of effective modes of the waveguide-interfaced topological atom array. In particular, this approach allows us to study the topological matter-light interaction in the scattering processes.

\subsection{Interfacing light and topological quantum matter via a waveguide}
As shown in Fig.~\ref{fig1}(a), we study a topological atom array coupled to photonic modes in a 1D waveguide with linear dispersion. The Hamiltonian of the waveguide is ($\hbar\equiv 1$)
\begin{equation}
H_{\mathrm{wg}} = \sum_{\alpha=r,l} \int dx \; \hat{a}_{\alpha}^{\dagger}(x) \left(\omega_0 -i  s_{\alpha} c \frac{\partial}{\partial x}\right)\hat{a}_{\alpha} (x),
\end{equation}
where $\hat{a}_l^{\dagger}$ ($\hat{a}_r^{\dagger}$) and $\hat{a}_l$ ($\hat{a}_r$) are respectively the creation and annihilation operators for the left (right) propagating photons; $s_{r}=+1, s_{l}=-1$ represent the right- and left-moving photons; $c$ is the photon velocity in the waveguide. The topological is described by the free energy $H_0=\sum_i \omega_0 \sigma_i^+\sigma_i^-$ and the SSH Hamiltonian~\cite{PhysRevLett.42.1698}
\begin{equation}
H_{\mathrm{ssh}} =  \left(J_{-} \sum_{i=\mathrm{odd}} \sigma_{i}^+ \sigma_{i+1}^- +  J_{+} \sum_{i=\mathrm{even}} \sigma_{i}^+ \sigma_{i+1}^-\right) + \mathrm{H.c.}, \label{Eqssh}
\end{equation}
where $\sigma_i^+=|e_i\rangle\langle g_i|$ depicts the transition from the ground state $|g_i\rangle$ to the excited state $|e_i\rangle$ of the $i$th atom, and the nearest neighbor flip-flop interactions change alternatively along the chain as $J_{\mp}=J_0(1\mp\cos\varphi)$. Here, $J_0$ and $\varphi$ are parameters that control the bandgap and localization of edge states. {In Fig.~\ref{fig1}(b), energy spectrum of the topological array with $5$ atoms is shown. For arrays with larger numbers of atoms, bulk states form upper and lower bands with a gap. The bandgap provides a large nonlinearity for the edge state. Different from the two edge modes in the SSH lattice with even number of sites~\cite{PhysRevResearch.2.012076}, only a single edge state exists at either the left ($0\leq\varphi<\pi/2$) or the right ($\pi/2<\varphi\leq \pi$) boundary of the topological array with an odd number of sites. Without loss of generality, we focus on the atom array with a left-localized edge state, i.e., $0\leq\varphi<\pi/2$. The wave functions of eigenmodes in the topological atom array are shown in Fig.~\ref{fig1}(c). Bulk states are extensive in all the atoms. However, the edge state populates odd sites due to topological protection. As the array becomes large, the edge state localizes to the left boundary.

In the Markovian approximation, the coupling between atoms and waveguide can be written as
\begin{equation}
H_{\mathrm{int}}=g \sum_{i,\alpha=r,l} \hat{a}_{\alpha}^{\dagger}(x_i)\sigma_i^- e^{-i s_{\alpha}k_0 x_i} + \mathrm{H.c.}, \label{Eqint}
\end{equation}
which is determined by the coupling strength $g$, the wave vector $k_0=\omega_0/c$, and the positions of atoms $x_i$. Therefore, the whole Hamiltonian becomes
\begin{equation}
H=H_{\mathrm{wg}}+H_0+H_{\mathrm{ssh}}+H_{\mathrm{int}} + H_{\mathrm{en}}, \label{Hall}
\end{equation}
where $H_{\mathrm{en}}$ represents the Hamiltonian of the environment.

\subsection{Multi-channel photon scattering}
Let us now study the photon scattering by the topological atom array. In the interaction picture, the Hamiltonian Eq.~(\ref{Hall}) can be written as
\begin{equation}
H(t)=H_{\mathrm{ssh}}+H_{\mathrm{int}}(t)+H_{\mathrm{en}},
\end{equation}
with $H_{\mathrm{int}}(t)=e^{i H_{\mathrm{wg}}t}H_{\mathrm{int}}e^{-i H_{\mathrm{wg}}t}$. For single-photon scattering, the input and output states for waveguide are, respectively, $|\psi_{\mathrm{in}}\rangle_{\mathrm{w}}=\exp(is_{\alpha}ckt_i)|1_{k\alpha}\rangle$ and $|\psi_{\mathrm{out}}\rangle_{\mathrm{w}}=\exp(is_{\beta}ckt_f)|1_{k\beta}\rangle$, where $k$ is the momentum of the photon; $\alpha,\beta=l,r$ label the propagation directions of the input and output photons; $t_i$ and $t_f$ are the initial and final times of the scattering process. The atom array is assumed to be in the ground state $|G\rangle$. Therefore, the scattering process can be formulated as the transition between input and output states
\begin{equation}
\mathcal{A}(T) = \langle G| \langle b_{\mathrm{out}}| U(T) |b_{\mathrm{in}}\rangle |G\rangle, \label{TranAmp}
\end{equation}
where the input (output) state for the environment and the waveguide is $|b_{\mathrm{in}(\mathrm{out})}\rangle=|\emptyset\rangle_{\mathrm{en}}|\psi_{\mathrm{in}(\mathrm{out})}\rangle_{\mathrm{w}}$, with $|\emptyset\rangle_{\mathrm{en}}$ being the vacuum state of the environment. The time evolution operator is $U(T)=\mathcal{T}\exp[-i\int_{t_i}^{t_i+T} dt H(t)]$ with $T=t_f-t_i$, and $\mathcal{T}$ being the time-ordering operator. In terms of the representation of unnormalized coherent states $|J_{k\alpha}\rangle=\sum_{n_k}J_{k\alpha}^{n_k}|n_{k\alpha}\rangle/\sqrt{n_{k\alpha}!}$, the single-photon states of the waveguide can be rewritten as~\cite{PhysRevA.92.053834}
\begin{eqnarray}
|1_{k\alpha}\rangle &=&\lim_{{J_{k\alpha}}\rightarrow 0}\frac{\delta}{\delta J_{k\alpha}}|\{J_{k\alpha}\}\rangle, \\
|1_{k\beta}\rangle &=&\lim_{{J_{k\beta}}\rightarrow 0}\frac{\delta}{\delta J_{k\beta}}|\{J_{k\beta}\}\rangle.
\end{eqnarray}
Expressing coherent states in terms of displaced vacuum states, we obtain the transition amplitude Eq.~(\ref{TranAmp})
\begin{equation}
\mathcal{A}(T) = \lim_{J_{k\alpha(\beta)}\rightarrow 0}\left(\frac{\delta}{\delta J_{k\beta}}\right)^{\ast}\frac{\delta}{\delta J_{k\alpha}} \mathcal{A}_J(T), \label{Amp}
\end{equation}
where
\begin{equation}
\mathcal{A}_J(T)=\mathcal{M}_J(T)\langle G|_b\langle \{0_{k\beta}\}| U_J(T) |\{0_{k\alpha}\}\rangle_b|G\rangle,
\end{equation}
with $\mathcal{M}_J(T)=\exp(|J_{k\alpha}|^2 e^{-i s_{\alpha} ck T}+|J_{k\beta}|^2 e^{-i s_{\beta} ck T})$ and the vacuum states of the environment and waveguide $|\{0_{k\alpha/k\beta}\}\rangle_b=|\emptyset\rangle_{\mathrm{en}} |\{0_{k\alpha/k\beta}\}\rangle$. Comparing with $U(T)$, the displaced time evolution operator $U_J(T)=\mathcal{T}\exp{-i \int_{t_i}^{t_i+T} dt\;[H(t) + H_d(t)]}$ contains an effective driving term
\begin{equation}
H_d(t)=\sum_{k} J_{k,\beta}^{\ast}\sigma^-_{k,s_{\beta}} e^{-is_{\beta}ck(t_f-t)} + J_{k,\alpha}\sigma^+_{k,s_{\alpha}} e^{-is_{\alpha}ck(t-t_i)}
\end{equation}
with $\sigma^-_{k,\pm}=(1/\sqrt{N})\sum_i g \sigma_i^- \exp[-i(k\pm k_0)x_i]$. After tracing out degrees of freedom of environment and waveguide, we obtain
\begin{equation}
\mathcal{A}_J(T)=\mathcal{M}_J(T) \langle G|\mathcal{T}e^{-i \int_{t_i}^{t_i+T} dt [H_{\mathrm{eff}} + H_d(t)]} |G\rangle, \label{AJ}
\end{equation}
where $H_{\mathrm{eff}}$ is the effective Hamiltonian. By taking the functional derivatives to Eq.~(\ref{AJ}), we apply the quantum regression theorem to obtain the single-photon transmission and reflection amplitudes~\cite{PhysRevA.92.053834}
\begin{eqnarray}
t&=&1-i\Gamma \sum_{i,j} G_{ij} \exp[i s_{\alpha} k_0 (-x_i+x_j)], \\
r&=&-i\Gamma \sum_{i,j} G_{ij} \exp[i s_{\alpha}k_0 (x_i + x_j)] ,
\end{eqnarray}
where $\Gamma=g^2/c$ denotes the spontaneous emission rate to the waveguide and $G_{ij}$ are matrix elements of the Green's function
\begin{equation}
G=\frac{1}{\delta\omega-H_{\mathrm{eff}}}.
\end{equation}
Here, $\delta\omega=ck$ is the frequency difference between the incident photon and atoms, and the effective Hamiltonian becomes
\begin{equation}
H_{\mathrm{eff}}=H_{\mathrm{ssh}}+H'_{\mathrm{en}} + H'_{\mathrm{wg}}, \label{effH}
\end{equation}
under the Markovian approximation. Specifically,
\begin{equation}
H'_{\mathrm{en}}=-i \Gamma_0\sum_i \sigma_i^+ \sigma_i^-,
\end{equation}
with $\Gamma_0$ being the decay rate to environment, and
\begin{equation}
H'_{\mathrm{wg}}=-i \Gamma \sum_{i,j} e^{i k_0 |x_i-x_j|} \sigma_i^+ \sigma_j^-.
\end{equation}
The Hamiltonian $H'_{\mathrm{wg}}$ represents the long-range interaction from the light-atom coupling in Eq.~(\ref{Eqint})~\cite{PhysRevLett.106.020501,caneva2015quantum}. It is clear that $H_{\mathrm{eff}}$ is non-Hermitian. The interplay of the coherent dynamics governed by the SSH Hamiltonian $H_{\mathrm{ssh}}$ and the incoherent interaction $H'_{\mathrm{wg}}$ determines the photon transport.

Compared with the optical responses~\cite{PhysRevA.92.053834,caneva2015quantum} of the nontopological atom array \emph{without} direct interaction, i.e., $J_0=0$, the topological atom array has a profound influence on the photon transport. In terms of eigenmodes of the waveguide-interfaced topological atom array, the amplitudes for the photon transmission and reflection are respectively
\begin{eqnarray}
t(\delta\omega)&=&1-i \Gamma\sum_j \frac{\bm{V}^{\dagger} |\psi_j^R\rangle  \langle \psi_j^L| \bm{V}}{\delta\omega-\Delta_j+i \Gamma_j}, \label{Eqtarray} \\
r(\delta\omega)&=&-i \Gamma\sum_j \frac{\bm{V}^{T} |\psi_j^R\rangle  \langle \psi_j^L| \bm{V}}{\delta\omega-\Delta_j+i \Gamma_j}, \label{Eqrarray}
\end{eqnarray}
where $\bm{V}=(e^{\pm i k_0 x_1}, e^{\pm i k_0 x_2}, \cdots)^{T}$ for the left- and right-incident photons, respectively; and the right and left eigenvectors $|\psi_j^R\rangle$ and $|\psi_j^L\rangle$ of $H_{\mathrm{eff}}$ in Eq.~(\ref{effH}) form the biorthogonal basis, i.e., $\langle \psi_j^L|\psi_{j'}^R\rangle=\delta_{jj'}$~\cite{brody2013biorthogonal}. The real and imaginary parts of $E$, i.e., $\Delta_j$ and $\Gamma_j=-\mathrm{Im}(E_j)$, denote the energy shift and the effective decay of the $j$th mode in $H_{\mathrm{eff}}$, respectively. And $\Gamma_j = \Gamma_0 + \tilde{\Gamma}_j$, where $\tilde{\Gamma}_j$ denotes the collective decay induced by the dissipative interaction $H'_{\mathrm{w}}$. The numerators in Eqs.~(\ref{Eqtarray}) and (\ref{Eqrarray}) characterize the changes of photonic states produced by the eigenmodes in the transmission and reflection processes. In our system, the edge state is regarded as a topological invariant and affects photon transport via the numerators of Eqs.~(\ref{Eqtarray}) and (\ref{Eqrarray}). For this purpose, we are interested in the scattering properties of photons which are resonant with the edge state. \emph{The topological features of the array are imprinted in the spatial profile of the eigenvectors}, $|\psi_j^R\rangle$ and $\langle\psi_j^L|$, \emph{and the structure of the spectrum, all of which eventually determine the photon transport}.

Collective many-body states are essential for photon transport. In the unstructured atom arrays, i.e., without direct interaction between atoms, photon scattering is dominated by a superradiant state with decay rate $\propto N\Gamma$ at mirror configurations~\cite{chang2012cavity}. This superradiant state gives rise to large Bragg reflection via collective enhancement~\cite{PhysRevA.83.013825,chang2012cavity,PhysRevLett.117.133603,PhysRevLett.117.133604}. Besides, there are many subradiant states, which can be employed for photon storage~\cite{PhysRevX.7.031024}. In the structured atom arrays with direct interactions, the many-body states become more complex due to the interplay between the waveguide-mediated interaction and the interaction in the many-body system. To explore the advantage of topology in manipulating photon transport, we consider the strong topological regime, i.e., $J_0\gg \Gamma$. The edge and bulk states in the topological atom array provide distinctive scattering channels. The multi-channel scattering approach provides a convenient way for studying light-matter interaction in the photon transport process.

\subsection{Topologically protected edge-state channel}
In the strong topological regime, one can expect that the localized edge state survives. In Fig.~\ref{fig1}(d), we show the real part $\Delta=\mathrm{Re}(E)$ of the energy spectrum $E$ of $H_{\mathrm{eff}}$ for $J_0\gg\Gamma$. The spectrum has the periodicity $\lambda_0=2\pi/k_0$ in $d$. As atomic spacing $d$ varies from $0$ to $\lambda_0$, the spectrum of bulk states is significantly changed due to the long-range interaction mediated by waveguide photons~\cite{solano2017super}. However, since the edge state is topologically protected from the bulk modes by the bandgap, it is only slightly shifted. In particular, both the left and right halves of the spectrum have the rotational symmetry by $\pi$ with respect to $(d,\Delta)=(\lambda_0/4,0)$ and $(3\lambda_0/4, 0)$, respectively. In other words, at $d=\lambda_0/4$ and $d=3\lambda_0/4$, the edge state has no energy shift from the coupling to waveguide. This protection comes from the chiral symmetry in the SSH atom array, which leads to polarization of the edge state. The wave function of left edge state is
\begin{equation}
|\alpha_{j_0}\rangle=\frac{1}{\sqrt{\mathcal{N}}}\sum_{i=\mathrm{odd}}(-1)^{\frac{i-1}{2}}\left(\tan\frac{\varphi}{2}\right)^{i-1}|i\rangle,
\end{equation}
where $j_0$ denotes the edge state and $\mathcal{N}$ is the normalization factor. Here, $|i\rangle=\sigma_{i}^+|G\rangle$, with $|G\rangle$ being the ground state of the atom array. Therefore, the average of effective Hamiltonian for the edge state is
\begin{equation}
\langle \alpha_{j_0}|H_{\mathrm{eff}}|\alpha_{j_0}\rangle = \langle \alpha_{j_0}|H_{\mathrm{wg}}'|\alpha_{j_0}\rangle - i \Gamma_0.
\end{equation}
At $d=\lambda_0/4$ and $d=3\lambda_0/4$, the coherent part of $H_{\mathrm{wg}}'$ only contains the coupling between odd- and even-site atoms. Because of the odd-site polarization of the edge state, we can have
\begin{equation}
\mathrm{Re}[\langle \alpha_{j_0}|H_{\mathrm{wg}}'|\alpha_{j_0}\rangle]=0.
\end{equation}
Namely, the edge state is topologically protected at $d=\lambda_0/4$ and $d=3\lambda_0/4$. At $d=0, \lambda_0/2$ and $\lambda_0$, the coherent part in $H_{\mathrm{wg}}'$ is vanishing, giving rise to zero shift. In other spacings, the edge state is not protected from the waveguide-mediated interaction and has energy shift. In Fig.~\ref{fig1}(e), we show the transmission spectra for $d=\lambda_0/4, \lambda_0/2$, and $3\lambda_0/4$. The different linewidths of the transmission at $\omega=0$ are determined by the decay rate of the edge mode. Dips appear in the transmission spectra when the incident photon is resonant with the bulk states of frequency around $\pm 2J_0$.

The bandgap in the topological atom array induces a large nonlinearity for the edge state, as shown in Fig.~\ref{fig2}(a). As a consequence, the topological atom array can be modeled as a superatom~\cite{PhysRevLett.124.023603} with nontrivial edge and bulk states. Such nonlinearity is important for probing and manipulating edge states~\cite{PhysRevLett.123.080501,PhysRevLett.124.023603,besedin2020topological,PhysRevResearch.2.012076}. Here we find that the bandgap is crucial for controlling the photon transport in the waveguide. The multi-channel formulas for photon transport in a waveguide Eqs.~(\ref{Eqtarray}) and (\ref{Eqrarray}) show that \emph{a photon can be scattered by bulk states even though it is resonant with the edge state}. It has been shown that there are subradiant many-body states in the 1D atom array which is coupled to a waveguide~\cite{PhysRevX.7.031024,PhysRevLett.122.203605}. The \emph{subradiance} means weak couplings to photonic modes in the waveguide. Therefore, the bandgap can get rid of the influence of these subradiant states in scattering processes if the frequency of the incident photon is in the bandgap, as shown in Fig.~\ref{fig2}(a). However, the superradiant bulk states are involved in the scattering when the bandgap is not large enough. Therefore, the bandgap provides a way to tune the photon scattering by balancing the waves scattered from edge and bulk states.

\begin{figure}[t]
\includegraphics[width=8cm]{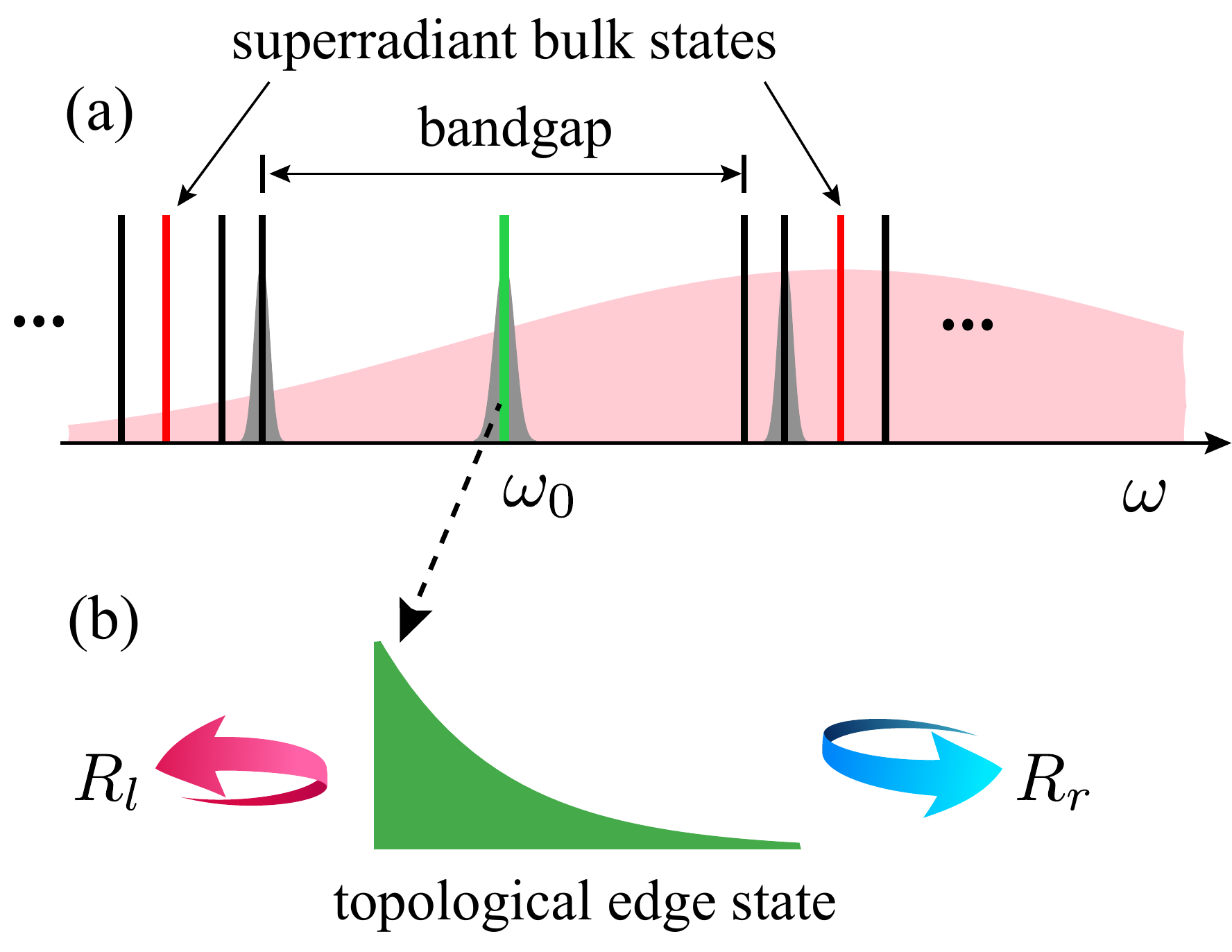}
\caption{(a) Multi-channel scattering for the topological atom array. In the waveguide-interfaced topological atom array, there are a few bulk states with huge decay rates. These superradiant bulk states are also important for the photon scattering. The purple and grey shapes denote the linewidths of the corresponding eigenmodes. (b) Anisotropic optical response via the channel of the topological edge state.}\label{fig2}
\end{figure}

\subsection{Photon detector: qubit vs topological superatom}
To understand and utilize the multi-channel scattering theory, we go back to the scattering problem of a single particle. For a single qubit, the amplitudes of transmission and reflection are described by~\cite{PhysRevLett.95.213001,shen2005coherent,chang2007single}
\begin{eqnarray}
t_1(\delta\omega)&=& 1- i \Gamma\frac{1}{\delta\omega + i (\Gamma+\Gamma_0)}, \label{Eqtqubit} \\
r_1(\delta\omega)&=& -i \Gamma\frac{1}{\delta\omega + i (\Gamma+\Gamma_0)}, \label{Eqrqubit}
\end{eqnarray}
with $r_1=t_1-1$. In single-qubit photon scattering, the detuning and decay rates determine the scattering processes. For example, at resonant driving $\delta \omega =0$, the photon is almost reflected for large Purcell factor $\Gamma/\Gamma_0$. The single-qubit controlled photon transport is useful for quantum control and information processing~\cite{PhysRevA.80.062109,PhysRevA.84.045801,peropadre2013scattering,PhysRevLett.122.173603}. A single qubit can also be used as a photon detector. However, detection efficiency $\eta_1=1-|t_1|^2-|r_1|^2$ of a single-qubit detector has an upper limit $50\%$. In order to improve the detection efficiency, various approaches have been suggested, e.g., using a three-level atom~\cite{PhysRevLett.100.093603,PhysRevLett.102.173602}, which has been realized in experiments~\cite{PhysRevLett.107.217401,inomata2016single} with the detection efficiency $60\% \sim  70\%$. To achieve reliable signal measurement for quantum computation, high-efficiency detection is required.

The topological atom array, or topological superatom formed by the edge and bulk states, shows a new strategy for quantum detection. For the single-qubit scattering, transmission and reflection processes are the same. In the multi-channel scenario, transmission and reflection processes are different because of the effective couplings between output photons and many-body states.  In the topological atom array, edge and bulk states provide separated scattering channels because of their distinctive properties in light-matter interaction~\cite{PhysRevResearch.2.012076}. In particular, the edge state plays an important role due to its topological features. In the following, we present the mechanism for single-photon detection via topology-enhanced optical nonreciprocity.

\begin{figure*}[t]
\includegraphics[width=18cm]{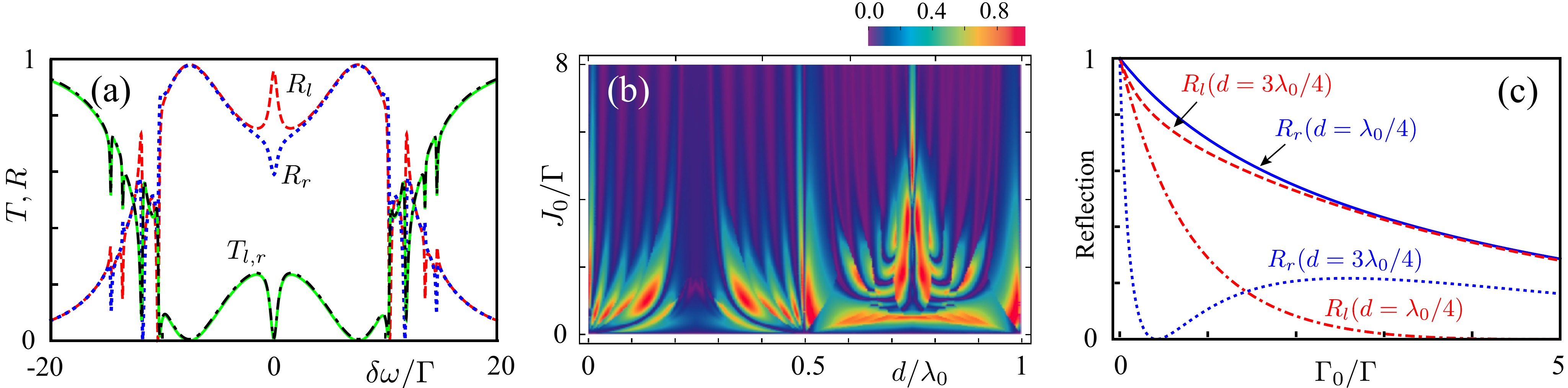}
\caption{(a) Transmission and reflection of a single photon through a topological atom array. Red-dashed (black-dot-dashed) and blue-dotted (green-solid) curves denote the reflection (transmission) for left- and right-incident photons, respectively. Reflections from the left and right are different at resonance. (b) Nonreciprocity changes with atomic spacing $d$ and coupling parameter $J_0$. When $J_0$ is zero, there is no nonreciprocity. When $J_0$ is large, the nonreciprocity has a maximum at $d=3\lambda_0/4$. (c) Reflection for left- and right-incident photons, represented by $R_l$ and $R_r$, at atomic spacings $d=\lambda_0/4$ and $d=3\lambda_0/4$. We consider $\Gamma_0/\Gamma=0.05$ in (a,b), $J_0/\Gamma=8$ in (a,c), $d=3\lambda_0/4$ in (a), $\varphi=0.3\pi$, atom number $N=11$ in (a,b,c).}\label{fig3}
\end{figure*}

\section{Nonreciprocal photon transport}\label{sec3}
Photon scattering is related to intrinsic properties of a quantum many-body system. In this section, we show how topological degrees of freedom of the atom array nontrivially affect photon transport in the waveguide.

\subsection{Topological edge state and anisotropy in the optical response}
The topological edge state is localized at the boundary of the system. This localization is a distinction between edge and bulk states. Depending on the degree of localization, the edge state may have different couplings to the waveguide, and decay rates. For the reason that we consider a single edge state in the system, the edge-state channel does not have inversion symmetry. The inversion symmetry breaking introduces anisotropy in the optical response, as schematically shown in Fig.~\ref{fig2}(b). The anisotropy is found in the photon reflection, i.e., nonreciprocal reflection. In Fig.~\ref{fig3}(a), we show the transmission and reflection spectra for the left- and right-incident photons, where the transmission is reciprocal. Here, $R_{\alpha}(\delta\omega)=|r_{\alpha}(\delta\omega)|^2$ and $T_{\alpha}(\delta\omega)=|t_{\alpha}(\delta\omega)|^2$, where $\alpha=r, l$ represents the right or left incident direction of the input photon. However, reflections for the left- and right-incident photons are different, when the incident photon is resonant with the edge state.

In Fig.~\ref{fig3}(a), we can also find nonreciprocal reflection when the incident photon is resonant with bulk states. Indeed, the nonreciprocal reflection can be observed in atom arrays with broken inversion symmetry, as we study below. However, it should be noted that topology plays a unique role in enhancing the nonreciprocity.

\subsection{Special spacing}
We now study the optical response of the topological atom array. To characterize the anisotropic feature, we define the reflection nonreciprocity
\begin{equation}
\Delta R=|R_l(0)-R_r(0)|,
\end{equation}
where the incident photon is assumed to resonate with the edge state. In the waveguide-interfaced topological atom array, the edge state at spacings $d=\lambda_0/4$ and $d=3\lambda_0/4$ is not shifted, i.e., $\Delta=0$ as shown in Fig.~\ref{fig1}(d). This means that at these two spacings the edge state is exactly protected. By increasing the topological nonlinearity (by enlarging the bandgap), the nonreciprocity behaves differently at these two spacings. In our model, the bandgap ($\propto J_0$) controls the nonlinearity [see Fig.~\ref{fig1}(b)].

Figure~\ref{fig3}(b) shows $\Delta R$ versus the spacing $d$ and the interaction parameter $J_0$. As expected, for vanishing $J_0$, the reflection is reciprocal. As $J_0$ increases, the position of the spacing $d$ at which the maximal nonreciprocity appears changes accordingly. For relatively large $J_0$, the maximal nonreciprocity appears at $d=3\lambda_0/4$, which we refer as a ``special spacing". The nonreciprocity induced by large $J_0$ at $d=3\lambda_0/4$ uncovers the nontrivial role of topology in altering the optical response. However, the nonreciprocity at another topology-protected spacing $d=\lambda_0/4$ is very small and is hardly changed by $J_0$. In Fig.~\ref{fig3}(c), we compare the reflections for $d=\lambda_0/4$ and $d=3\lambda_0/4$ versus $\Gamma_0/\Gamma$. The environment-induced decay $\Gamma_0$ alters the nonreciprocity at these two spacings. Large nonreciprocity can be realized with a tiny $\Gamma_0/\Gamma$ for $d=3\lambda_0/4$. At $d=\lambda_0/4$, the nonreciprocity is small for tiny $\Gamma_0/\Gamma$. But, it can be increased when $\Gamma_0/\Gamma$ grows. By considering different values of $\Gamma_0/\Gamma$, the largest nonreciprocity is realized at $d=3\lambda_0/4$.

\subsection{Reflection nonreciprocity versus broken time-reversal and inversion symmetries}
The time-reversal symmetry of the system (topological atom array+waveguide) is broken by $\Gamma_0$. As shown by the transmission and reflection versus $\Gamma_0$ in Fig.~\ref{fig4}(a), the reciprocity of the reflection for $\Gamma_0=0$ results from time-reversal symmetry. When $\Gamma_0$ increases, in contrast to the almost unchanged reflection of the left-incident photon, the reflection of the right-incident photon exhibits non-monotonic behavior, which reaches its minimum at $\Gamma_{0m}$. For the right-incident photon, the decrease of reflection is much faster than the increase of transmission. The nonconserved photon number suggests that the photon is lost to the environment. A surprising fact is that the photon loss is quite large even with small $\Gamma_0/\Gamma$. In addition, the transmissions of left- and right-incident photons are reciprocal and slightly changed as $\Gamma_0$ varies.

The enhanced nonreciprocity is produced by the broken inversion symmetry}. In Fig.~\ref{fig4}(b), $\Gamma_{0m}$ versus $\varphi$ is plotted for $N=11$ and $21$. For $0\leq\varphi<\pi/2$, the edge mode appears on the left boundary. Therefore, the inversion symmetry is broken. When $\varphi$ is equal to $\pi/2$, the inversion symmetry is restored. As $\varphi$ approaches $\pi/2$, $\Gamma_{0m}$ increases very fast. We can infer from Fig.~\ref{fig4}(a) that the reflection nonreciprocity disappears for $\varphi \rightarrow \pi/2$. When $\varphi$ is away from $\pi/2$, $\Gamma_{0m}$ is reduced to very small values. This means that at these values, a large reflection nonreciprocity can be obtained (we assume $\Gamma_0/\Gamma$ to be small). The nonreciprocity versus $\varphi$ is shown in the inset of Fig.~\ref{fig4}(b). The red-dashed and blue-solid curves correspond to $N=11$ and $N=21$, respectively. As expected, the nonreciprocity reduces to zero as $\varphi$ is increased to $\pi/2$. The increase of $N$ can further reduce $\Gamma_{0m}$. As a consequence, a tiny environment-induced decay is able to yield a huge nonreciprocity.

\begin{figure}[t]
\includegraphics[width=8.5cm]{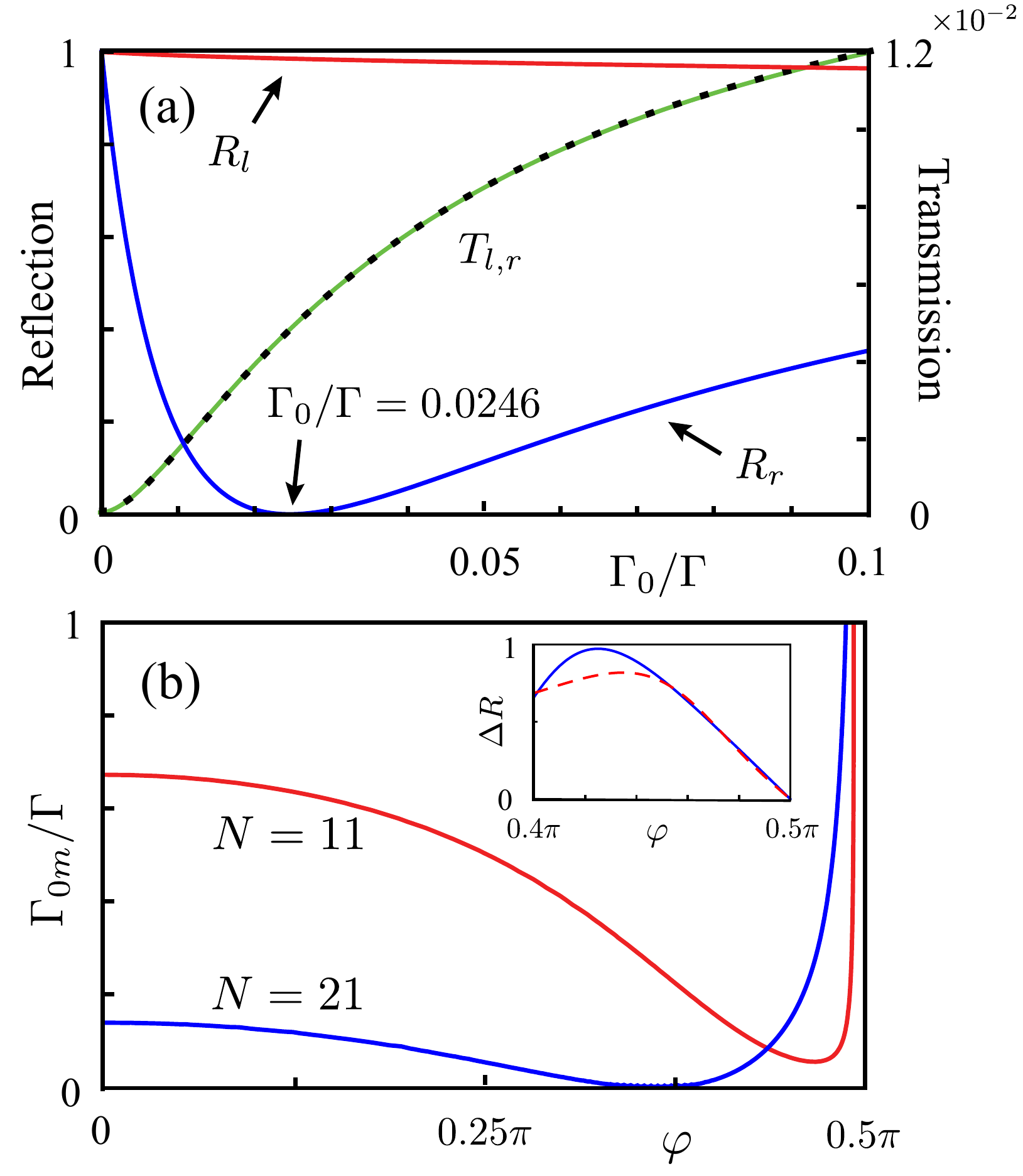}
\caption{(a) Effect of broken time-reversal symmetry. Reflections and transmissions for left- and right-incident photons change with $\Gamma_0$. The reflection $R_r$ (blue-solid) for right-incident photon is sensitive to $\Gamma_0$ and reduces to zero when $\Gamma_0/\Gamma\simeq0.0246$. The transmissions for left- and right-incident photons, denoted respectively by green-solid and black-dotted curves, are the same. (b) Broken inversion symmetry ($0<\varphi<\pi/2$). Parameter $\Gamma_{0m}$, defined by $R_r(\Gamma=\Gamma_{0m})=0$, versus $\varphi$. The inset shows the nonreciprocity versus $\varphi$ for $\Gamma_0/\Gamma=0.05$. The red-dashed and blue-solid curves correspond to $N=11$ and $N=21$, respectively. We consider $\varphi=0.3\pi, N=21$ in (a), and $J_0/\Gamma=8, d=3\lambda_0/4$ in (a,b).}\label{fig4}
\end{figure}

\section{Topological matter-light interaction}\label{sec4}
To understand the topology-enhanced reflection nonreciprocity found above, we need to study topological matter-light interactions in the scattering processes. By means of the multi-channel scattering formulas, in the following we elucidate the distinctive roles played by edge state and bulk states in the exotic optical response.

\subsection{Interaction spectra}
The multi-channel scattering formulas Eqs.~(\ref{Eqtarray}) and (\ref{Eqrarray}) show that photon propagation is influenced by the many-body states of the waveguide-interfaced atom array. To study the light-matter interaction in the photon scattering, we define the interaction spectra
\begin{eqnarray}
\Xi_j&=&\bm{V}^T |\psi_j^R\rangle  \langle \psi_j^L| \bm{V}, \label{EqXi1} \\
\tilde{\Xi}_j&=&\bm{V}^{\dagger} |\psi_j^R\rangle  \langle \psi_j^L| \bm{V}, \label{EqXi2}
\end{eqnarray}
i.e., the numerators in Eqs.~(\ref{Eqtarray}) and (\ref{Eqrarray}), which show the transformations of the photon states $\bm{k} \rightarrow -\bm{k}$ and $\bm{k} \rightarrow \bm{k}$ in the reflection and transmission processes. The interaction spectra can be understood as overlaps of propagating photon modes and eigenmodes of the effective Hamiltonian in the scattering processes. The symmetry-protected interaction in the topological atom array introduces nontrivial many-body states, including the topological edge state, with real wave functions. However, by considering the waveguide-mediated non-Hermitian Hamiltonian, the effective modes have complex amplitudes in their wave functions.

\begin{figure*}[t]
\includegraphics[width=18cm]{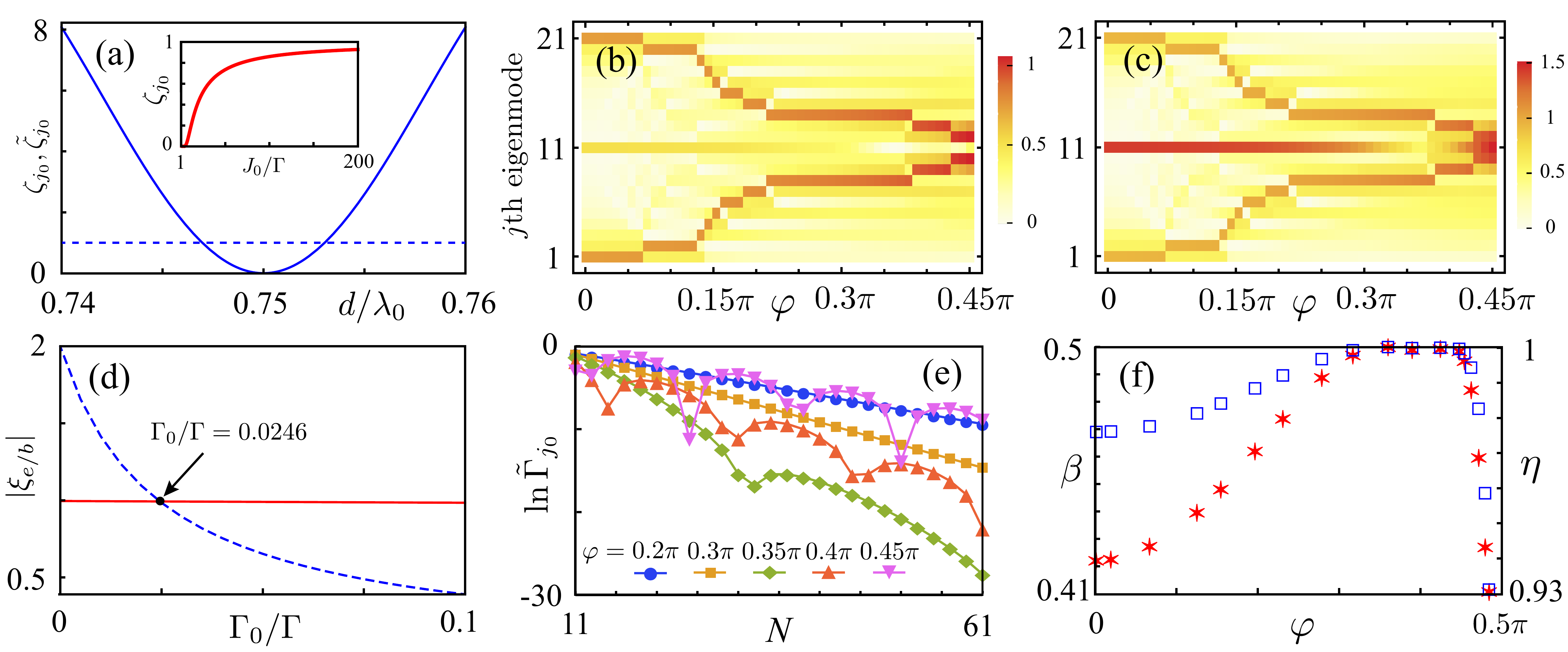}
\caption{(a) Relative strength for edge state reflecting photons coming from the left and right directions. Here, we define $\zeta_{j_0}=|\Xi_{j_0}^l/\Xi_{j_0}^r|$ and $\tilde{\zeta}_{j_0}=|\tilde{\Xi}_{j_0}^l/\tilde{\Xi}_{j_0}^r|$. The blue-solid and blue-dashed curves respectively represent $\zeta_{j_0}$ and $\tilde{\zeta}_{j_0}$. The scattering process of transmission is found to be reciprocal. However, the reflection process is nonreciprocal. For given parameters, $\zeta_{j_0}$ has a minimum at $d=3\lambda_0/4$. (b) and (c) show $|\xi_j|$ ($j=1,2,\cdots,N$) for photons coming from the left and right, respectively. The vertical axis labels the eigenmodes of $H_{\mathrm{eff}}$. (d) Absolute values of $\xi_{e}=\xi_{j_0}$ (blue-dashed) and $\xi_{b}=\sum_{j\neq j_0}\xi_j$ (red-solid), which correspond to the edge mode and bulk modes, for the right-incident photon. (e) The scaling behaviors between $\ln \tilde{\Gamma}_{j_0}$ and $N$ at different $\varphi$. (f) The $\beta$ factor of the edge state (blue square), and photon absorption $\eta$ (red star) for the right-incident photon at $\Gamma_0=\Gamma_{0m}$. We consider $J_0/\Gamma=8, d=3\lambda_0/4$ for (a,...,f), $\varphi=0.3\pi$ for (a,d), $N=21$ for (a,b,c,d,f), $\Gamma_0/\Gamma=0.05$ for (a,b,c).}\label{fig5}
\end{figure*}

It turns out that for vanishing direct interactions, i.e., $J_{0}=0$, $\Xi_j$ ($\tilde{\Xi}_j$) are the same for the left- and right-incident photons. At $\varphi\neq 0.5\pi$, nonzero $J_0$ makes $\Xi_j$ different for left- and right-incident photons. When $J_0$ is much larger than $\Gamma$, the edge state is separated from the bulk states. We can employ perturbation theory to obtain effective edge state. Therefore, the effective edge state can be approximately written as
\begin{equation}
|\psi_{j_0}^R\rangle \approx |\alpha_{j_0}\rangle + \sum_{j\neq j_0} \frac{\langle \alpha_j| H' |\alpha_{j_0}\rangle }{\varepsilon_{j_0} - \varepsilon_j}|\alpha_j\rangle, \label{EqedgeR}
\end{equation}
and
\begin{equation}
\langle\psi_{j_0}^L| \approx \langle\alpha_{j_0}| + \sum_{j\neq j_0} \frac{\langle \alpha_{j_0}| H' |\alpha_{j}\rangle }{\varepsilon_{j_0} - \varepsilon_j}\langle\alpha_j|, \label{EqedgeL}
\end{equation}
where $H'=H'_{\mathrm{en}} + H'_{\mathrm{wg}}$, and $\varepsilon_j$ are energies of $H_{\mathrm{ssh}}$ with eigenvectors $|\alpha_j\rangle$. As a consequence, the interaction spectrum of the edge-state channel in the reflection process can be written as
\begin{equation}
\Xi_{j_0}\approx \mathcal{X}_{j_0 j_0}-\sum_{j\neq j_0} (a_j \mathcal{X}_{j j_0} + b_j \mathcal{X}_{j_0 j} ), \label{EqXip}
\end{equation}
where $\mathcal{X}_{l k}=\langle \alpha_{l}| \bm{V} \bm{V}^T |\alpha_{k}\rangle $, $a_j=\langle \alpha_{j_0}| H'|\alpha_j\rangle/\varepsilon_{j}$ and $b_j=\langle \alpha_{j}| H'|\alpha_{j_0}\rangle/\varepsilon_{j}$. Similarly, for the transmission process,
\begin{equation}
\tilde{\Xi}_{j_0}\approx \mathcal{Y}_{j_0 j_0}-\sum_{j\neq j_0} (a_j \mathcal{Y}_{j j_0} + b_j \mathcal{Y}_{j_0 j} ), \label{EqXitp}
\end{equation}
with $\mathcal{Y}_{l k}=\langle \alpha_{l}| \bm{V}\bm{V}^{\dagger} |\alpha_{k}\rangle $. The reflection interaction spectrum for the photon with different incident direction of Eq.~(\ref{EqXip}) is $\Xi_{j_0}'=\mathcal{X}^{\ast}_{j_0 j_0}-\sum_{j\neq j_0} (a_j \mathcal{X}^{\ast}_{j j_0} + b_j \mathcal{X}^{\ast}_{j_0 j} )$. Because $H'$ is non-Hermitian,
\begin{equation}
|\Xi'_{j_0}|\neq |\Xi_{j_0}|.
\end{equation}
For the transmission, because $H'$ is a symmetric matrix, it can be shown that $\tilde{\Xi}'_{j_0}=\tilde{\Xi}_{j_0}$, where $\tilde{\Xi}'_{j_0}$ denotes the transmission interaction spectrum for the photon with different direction compared to $\tilde{\Xi}_{j_0}$. The isotropic transmission and anisotropic reflection can also be found for bulk-state channels by considering high-order perturbations. Namely, the interaction spectra for reflection and transmission are anisotropic and isotropic, respectively.

In Fig.~\ref{fig5}(a), we show the ratio between interaction spectra of reflection for photons coming from different directions
\begin{equation}
\zeta_{j_0}=\frac{|\Xi_{j_0}^l|}{|\Xi_{j_0}^r|},
\end{equation}
(blue-solid curve) as a function of $d$ at $J_0/\Gamma=8$, where $l$ ($r$) represents the left- (right-) incident photon. It is clear that $\zeta_{j_0}$ has a minimum at $d=3\lambda_0/4$, which implies a small overlap of the left-propagating photon and the edge state at $d=3\lambda_0/4$ during the reflection process. In other words, the left-incident photon barely couples to the effective edge state; however, the right-incident photon strongly couples to it. The blue-dashed line represents the ratio between interaction spectra of transmission for left- and right-incident photons
\begin{equation}
\tilde{\zeta}_{j_0}=\frac{|\tilde{\Xi}_{j_0}^l|}{|\tilde{\Xi}_{j_0}^r|}.
\end{equation}
Distinctive to the nonreciprocity in the reflection process, a reciprocal behavior is found for the transmitted photon. The inset of Fig.~\ref{fig5}(a) shows $\zeta_{j_0}$ at $d=3\lambda_0/4$ as $J_0$ is changed. When $J_0$ is large enough, $\zeta_{j_0}$ approaches one. This means that the scattering channel of the edge state loses the reflection nonreciprocity when bulk states are negligible in the system. Therefore, the bandgap controls the effective edge state and alters the nonreciprocal behavior of the edge-state scattering channel.

By taking account of detunings and decay rates, the contributions from eigenmodes to reflection can be characterized via
\begin{equation}
\xi_j=\frac{\Xi_j \Gamma }{-\Delta_j+i (\Gamma_0 + \tilde{\Gamma}_j)}.
\end{equation}
Because we assume that the edge state is resonantly driven, $\Delta_{j_0}=0$. In Figs.~\ref{fig5}(b) and \ref{fig5}(c), the absolute values of different components $\xi_j$ explicitly show the tiny and large contributions from the edge mode for the left- and right-incident photons, respectively. With a large bandgap, the effective edge state differentially scatters photons coming from different directions. Moreover, there are pairs of bulk states which have large scattering amplitudes and change with $\varphi$. Comparing with the effective edge state, these scattering channels of superradiant bulk states are less anisotropic. The distinctive optical properties of edge-state and bulk-state channels show the importance of topology in controlling photon transport.

The anisotropic interaction spectra for the reflection $\Xi_j$ are prerequisite for nonreciprocal reflection. Another essential condition is the environment-induced decay $\Gamma_0$. If $\Gamma_0$ is zero, the reflection is reciprocal, due to the conservation of the photon number. When $\Gamma_0$ is tiny, $\xi_{j_0}$ can be significantly changed because of the subradiance of the edge state. But those $\xi_{j}$ for bulk states are hardly changed by a small $\Gamma_0$ because of the bandgap. As a result, the nonreciprocal reflection of the photon is produced by considering all scattering channels. In other words, the nonreciprocity is produced by different roles of $\Gamma_0$ in changing the $\xi_j$ of resonant and nonresonant scattering channels. If the resonant channel is subradiant, large nonreciprocity can be realized by small $\Gamma_0/\Gamma$.

\subsection{Quenched reflection via destructive interference between waves reflected by edge and bulk modes}
For the right-incident photon, due to the finite coupling to the edge mode, the interference of waves reflected by the edge mode and bulk modes gives rise to the left outgoing wave. In Fig.~\ref{fig5}(d), absolute values of the contributions from the edge and bulk modes, i.e.,
\begin{equation}
\xi_e=\xi_{j_0}, \quad\quad \xi_b=\sum_{j\neq j_0}\xi_{j},
\end{equation}
are shown for the right-incident photon at $\varphi=0.3\pi$. For a closed system without $\Gamma_0$, the contributions from the edge mode and bulk modes are $\xi_{e}=2e^{i\phi _{0}}$ and $\xi_{b}=-e^{i\phi _{0}}$, respectively, with $\phi_0=\pi/2$. Note that there is a $\pi$ phase shift for the reflected waves from the edge and bulk modes. As a consequence, the photon is completely reflected.

When the environment-induced decay $\Gamma_0$ is turned on, the reflection $\xi_{b}\sim-e^{i\phi _{0}}$ from bulk states is hardly affected by the small $\Gamma_0$, since $|\Delta_{j\neq j_0}|$ or $\tilde{\Gamma}_{j\neq j_0}$ are much larger than $\Gamma_0$. However, because the edge mode has zero energy and a tiny decay rate $\tilde{\Gamma}_{j_0}$ at the special atomic spacing $d=3\lambda_0/4$, the small $\Gamma_0$ drastically reduces the reflection $\xi_{e}=e^{i\phi _{0}}$ from the edge mode by half when $\Gamma_0=\tilde{\Gamma}_{j_0}$, which induces a vanishing reflection
\begin{equation}
\xi_{e}+\xi_{b} \sim0,
\end{equation}
and the maximal nonreciprocity. Here, the tiny decay rate of the edge state is significant to achieve an enhanced nonreciprocity. We find that, for some values of $\varphi$, the collective decay rate for the effective edge state at $d=3\lambda_0/4$ exhibits a scaling,
\begin{equation}
\tilde{\Gamma}_{j_0}\sim \exp(-\nu N), \label{scaling}
\end{equation}
as shown in Fig.~\ref{fig5}(e). Due to this scaling behavior, $\Gamma_{0m}\sim\tilde{\Gamma}_{j_0}$ decreases as the size of the array increases, in good agreement with Fig.~\ref{fig4}(b). However, when $\varphi$ is close to $\pi/2$, such scaling behavior in Eq.~(\ref{scaling}) breaks down. The effective edge state may have a high decay rate for a large atom array. Therefore, a large $\Gamma_0$ is required to cancel the reflection. In this scenario, the transmission is increased [see Fig.~\ref{fig4}(a)], leading to reduced nonreciprocity.

\begin{figure}[b]
\includegraphics[width=8.5cm]{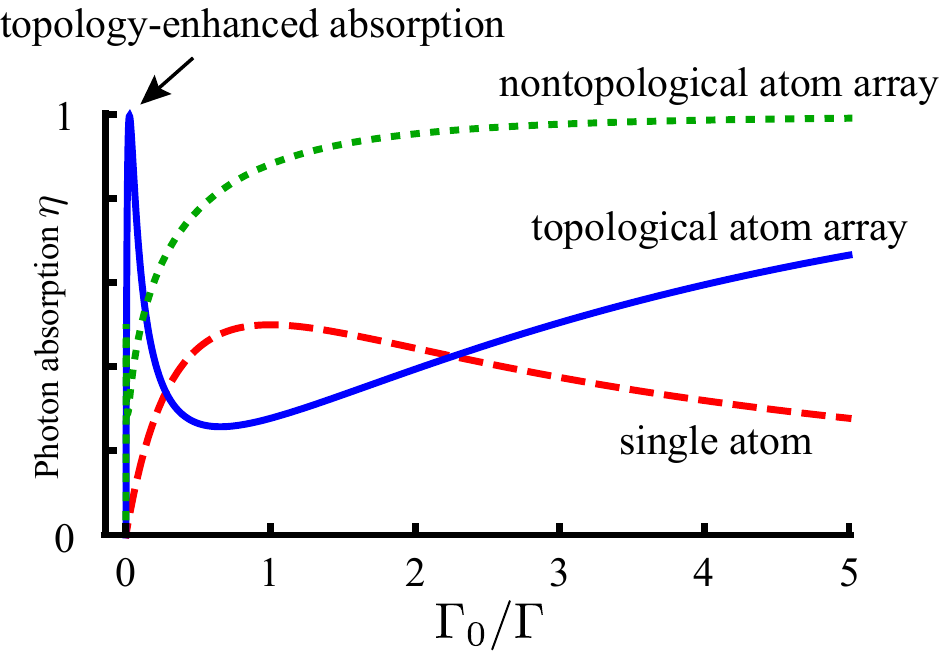}
\caption{Photon absorption from the waveguide. Blue-solid, red-dashed and green-dotted curves correspond to the topological atom array, single atom, and nontopological atom array without direct interaction, respectively. The topological atom array enhances photon absorption for $\Gamma_0/\Gamma\ll 1$. We consider $N=21, d=3\lambda_0/4$ for the atom array and topological atom array, $J_0/\Gamma=8,\varphi=0.3\pi$ for the topological atom array.}\label{fig6}
\end{figure}

\subsection{Topology-enhanced absorption of photon}\label{sec5}
Quantum scattering by edge and bulk modes yields an anomalous photon transport. We now use the beta factor~\cite{PhysRevB.75.205437,PhysRevLett.99.023902}
\begin{equation}
\beta=\frac{\tilde{\Gamma}_{j_0}}{\tilde{\Gamma}_{j_0}+\Gamma_{0}},
\end{equation}
to characterize the photon decay from the edge mode to the waveguide. When a single atom is coupled with the waveguide, a higher beta factor corresponds to larger photon emission to the waveguide~\cite{PhysRevLett.113.093603,PhysRevLett.115.153901}. However, our study shows that, due to the interference between reflective waves from edge and bulk modes, the photon in the waveguide totally emits to the environment even with a high beta factor, e.g., $\beta \sim1/2$, as presented in Fig.~\ref{fig5}(f). Here, the photon absorption from the waveguide is represented by
\begin{equation}
\eta=1-T_r-R_r,
\end{equation}
for a right-incident photon. As the beta factor becomes one half, the absorption efficiency is highest.

In Fig.~\ref{fig6}, we show the photon absorption for a single atom, an atom array without direction interaction, and also for topological atom array. The single atom has its largest photon absorption $\eta=0.5$ at $\Gamma_0/\Gamma=1$. The absorption can be enhanced by increasing the number of atoms in the array~\cite{PhysRevLett.102.173602}. The complete absorption can be realized when $\Gamma_0$ is much larger than $\Gamma$. However, the topological atom array enhances photon absorption for $\Gamma_0/\Gamma\ll 1$. The enhancement at this regime is useful photon detection in superconducting quantum circuits.

\section{Implementation and application}

\begin{figure}[b]
\includegraphics[width=8.5cm]{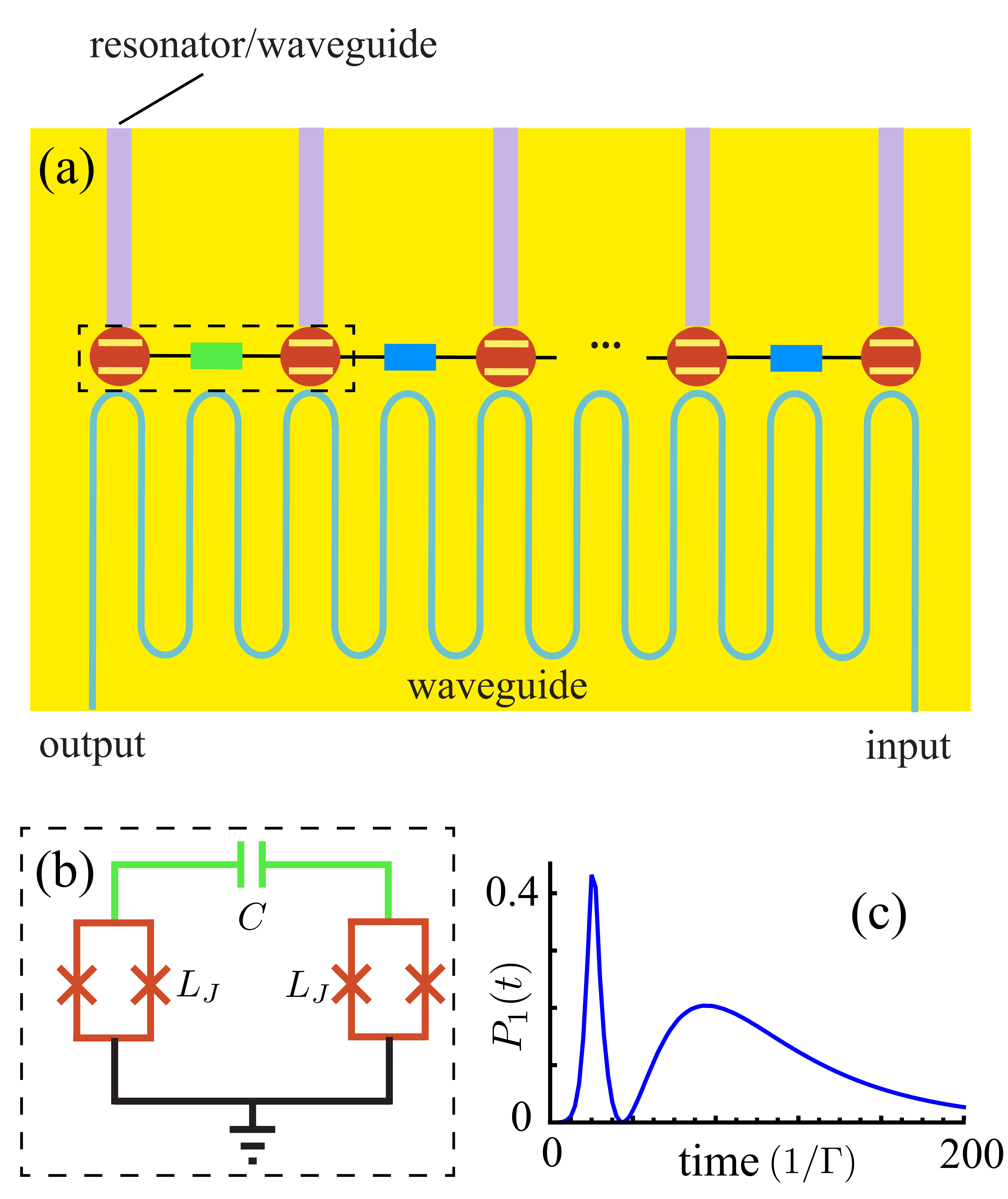}
\caption{(a) Waveguide QED with a topological atom array in superconducting quantum circuits. (b) Interaction between two superconducting artificial atoms. (c) Population dynamics of the first atom at the left boundary. The parameters used here are $d=3\lambda_0/4, J_0/\Gamma=8, \varphi=0.3\pi, N=21, \Gamma_0/\Gamma=0.0246$.}\label{fig7}
\end{figure}

\subsection{Scheme for superconducting quantum circuits}
Superconducting quantum circuits provide versatile interfaces for light-matter interactions~\cite{you2011atomic,gu2017microwave,carusotto2020photonic}, and have advanced controllability in atom-atom interaction and light-atom coupling~\cite{astafiev2010resonance,PhysRevLett.108.263601,peng2016tuneable,forn2017ultrastrong}. Topological atom arrays~\cite{PhysRevLett.123.080501,besedin2020topological} and waveguide-interfaced multi-atom systems~\cite{van2013photon,mirhosseini2019cavity,kannan2019waveguide,kannan2020generating,vadiraj2020engineering,Brehm2020waveguide} have been realized. Moreover, the topology-enhanced photon absorption at $\Gamma_0/\Gamma\ll 1$ is observable in superconducting quantum circuits, because of the strong coupling between superconducting artificial atoms and microwave waveguides. In particular, an extremely small $\Gamma_0/\Gamma$ has been demonstrated, e.g., $\Gamma_0/\Gamma \approx 0.005$ in Ref.~\cite{mirhosseini2019cavity}. These experimental achievements in superconducting quantum circuits make it promising to realize the waveguide-interfaced topological atom array, as shown in Fig.~\ref{fig7}(a). The interaction between superconducting artificial atoms can be realized via capacitors, as shown in Fig.~\ref{fig7}(b). In addition to superconducting quantum circuits, our model is feasible for other systems where the direct interaction can be realized, e.g, via dipole coupling~\cite{cheng2017waveguide,PhysRevLett.123.217401,mirza2020dimer}.

The scattering method studies the optical response of the system in the asymptotic limit, i.e., $T\rightarrow\infty$. However, during the scattering process, atoms in the array have a population dynamics~\cite{PhysRevA.92.053834}. Therefore, by coupling atoms with resonators, the photon absorption can be directly observed via measuring excitations of the atoms. Moreover, due to topological protection, the edge atom is mostly excited. Therefore, the detection of the edge atom simplifies the experimental measurement of the absorption of the incident photon. For a left-propagating single-photon wave package $f(k)=\sqrt{\gamma/\pi}e^{-i k x_N}/(k-i \gamma)$ with width $1/\gamma$, the dynamics of the edge atom is shown in Fig.~\ref{fig7}(c). As the incident photon interacts with the topological atom array, the population of the edge atom is increased and exhibits nontrivial dynamics. The population dynamics of the edge atom is related to the width $\gamma$ of the single photon. In Fig.~\ref{fig7}(c), we consider $\gamma/\Gamma=0.01$. The topology-enhanced absorption of light enables the detection of a weak signal.

In superconducting quantum circuits, the environment-induced decay can be made very small. In this case, one can replace the resonators with waveguides, and assume a weak coupling between atoms and additional waveguides. Therefore, the decay rate $\Gamma_0$ is mainly produced by the additional waveguide. As a consequence, the photon in the primary waveguide can be transferred to the additional waveguide coupled to the edge atom.

\begin{figure}[t]
\includegraphics[width=7.5cm]{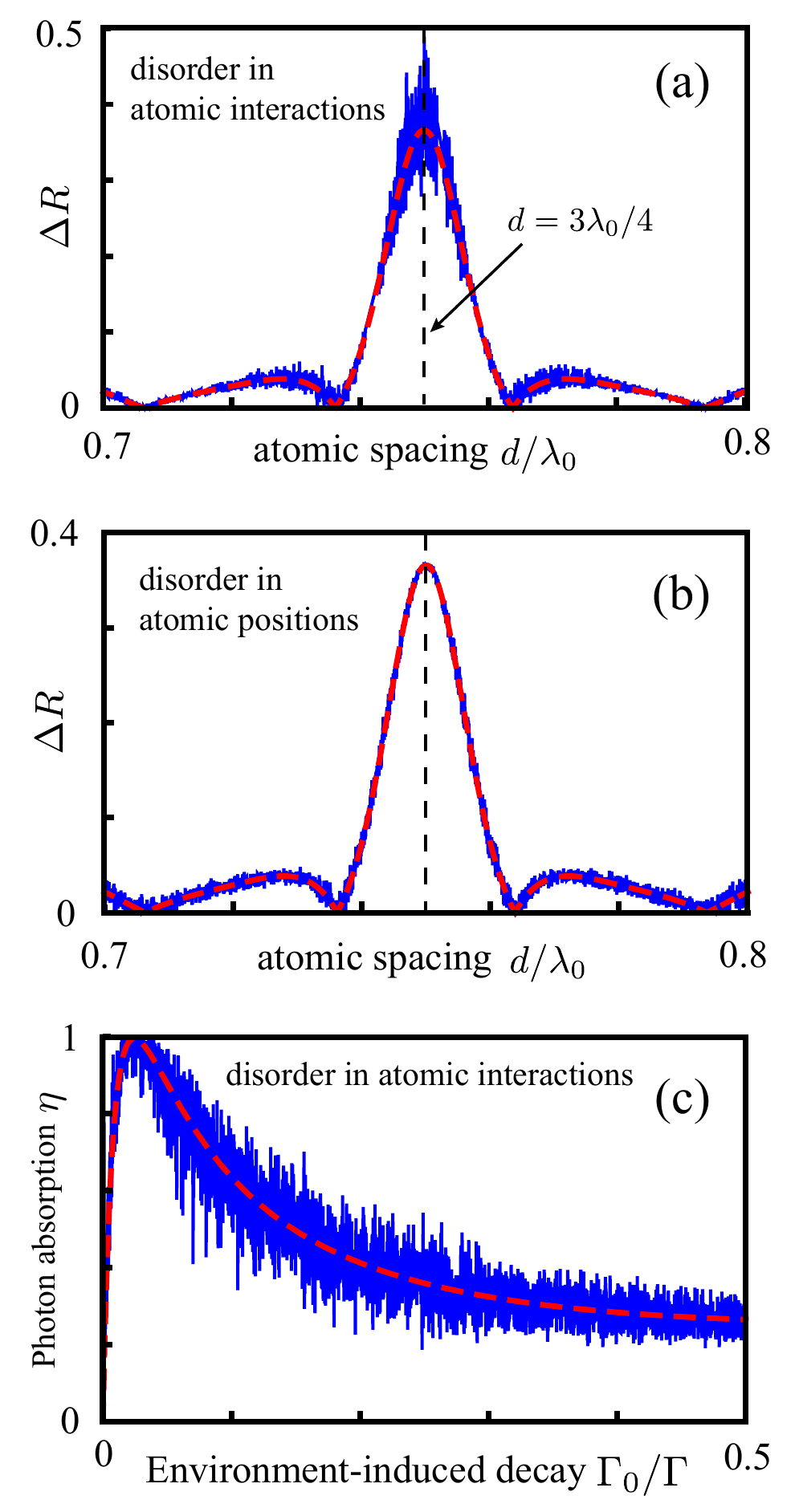}
\caption{(a) Reflectional nonreciprocity $\Delta R$ versus atomic spacing. The red-dashed and blue-solid curves correspond to zero and finite disorder in atomic interactions. (b) Reflectional nonreciprocity for zero (red-dashed) and finite (blue-solid) disorder in atomic positions. Note the peak in (a,b) when $d=3\lambda_0/4$. (c) Photon absorption $\eta$ versus $\Gamma_0/\Gamma$, where $\Gamma_0$ and $\Gamma$ denote atomic decays to environment and waveguide, respectively. The red-dashed and blue-solid curves correspond to zero and finite disorder in atomic interactions, respectively. Here, we consider $\varphi=0.3\pi, J_0/\Gamma=8$ in (a,b,c), $N=11, \Gamma_0/\Gamma=0.05$ in (a,b), $N=21$ in (c).}\label{fig8}
\end{figure}

\subsection{Effect of imperfections}
We have studied optical properties of the topological atom array. Let us now consider imperfections in the system. In Fig.~\ref{fig8}(a), we show the reflectional nonreciprocity of the topological atom array with disordered interactions $J_{i,i+1}+ \epsilon_{i}$ between the $i$th and $(i+1)$th atoms. Here, $\epsilon_{i}$ are randomly distributed $\epsilon_{i} \in [-\Gamma,\Gamma]$. Figure \ref{fig8}(a) shows how the disorder in atomic interactions affects the nonreciprocity. Note that around the special atomic spacing $d=3\lambda_0/4$, the nonreciprocity is more sensitive to the disorder, and far less sensitive elsewhere. Figure \ref{fig8}(b) shows how the nonreciprocity is changed by disorder in the atomic positions. For superconducting artificial atoms with frequency $\omega_0=2\pi\times 6$ GHz, the wavelength of photons is $\lambda_0=0.05$ m. The scale of superconducting artificial atoms is around $100$ $\mu$m. Therefore, we consider the position of $i$th atom $x_i=(i-1)d + \tau \lambda_0$ with disorder strength $\tau \in [-0.002, 0.002]$. We find that, near the special atomic spacing $d=3\lambda_0/4$, the reflectional nonreciprocity $\Delta R$ is robust to disorder in the position. Figure \ref{fig8}(c) shows the photon absorption for the disorder in atomic interactions $\epsilon_{i} \in [-\Gamma,\Gamma]$. The disorder affects the photon absorption, but not in a very negative manner. Note that a high photon absorption can be obtained for low values of the environment-induced decay $\Gamma_0$. Therefore, topology-enhanced photon absorption can still be obtained in systems with weak or moderate disorder.

\section{Discussions and Conclusions}
The superradiant and subradiant states are of great interest for novel optical phenomena and practical applications~\cite{scully2009super,PhysRevLett.115.243602}. Here, we pinpoint the interplay between superradiant and subradiant states in the photon scattering process with topological protection. Our results show the importance of the topological bandgap in manipulating photon transport. Due to the bandgap, subradiant bulk states are irrelevant in the optical response when the incident photon is resonant with the edge state. Therefore, photon transport through the topological atom array is {controlled by the subradiant edge state and superradiant bulk states}. Waves scattered by the edge and bulk states have destructive quantum interference, giving rise to zero reflection. Our study is based on the multi-channel scattering theory, in which the interaction spectra characterize light-matter interaction in the transport process and help us to understand the relation between the unconventional complete photon absorption and topology-protected many-body states.

Waveguide QED has a wide range of applications, including quantum computation, quantum network and quantum devices~\cite{lodahl2017chiral,RevModPhys.89.021001,RevModPhys.90.031002}. Photon detection plays a fundamental role in the waveguide QED. For a single-qubit photon detector, the detection efficiency has an upper limit of $50\%$. By increasing the number of qubits, the detection efficiency can be increased for low-coherence qubits, i.e., with large environment-induced decay. In superconducting quantum circuits, which are promising for quantum computation, this multi-qubit scheme is not efficient because of the long coherence times of superconducting qubits. We here show that the topological array is able to realize high-efficiency photon detection with long coherence qubits. We employ the topological protection of light-matter interaction in the waveguide. A small environment-induced decay $\Gamma_0$ can yield large nonreciprocal reflection due to the subradiant edge state with topological protection. Hence, the photon can be perfectly absorbed and detected. Our work presents an alternative way to realize complete photon absorption~\cite{PhysRevLett.105.053901,wan2011time}. The topological matter-light interaction is not only useful for photon detection as we study here, but also promising for topological quantum nondemolition measurement of Majorana qubits~\cite{PRXQuantum.1.020313}. Because of diverse types of topological matter, the topological matter-light interaction might be studied in other systems, e.g., with multi-atom interactions~\cite{PhysRevResearch.2.013135}. Recently, the interfaces between topological waveguides and atoms have been investigated~\cite{bello2019unconventional,kim2020quantum,leonforte2020vacancy}, enabling the study of topological light-matter interaction.

In conclusion, in this work, we study topology-protected light-matter interaction and show the potential of topological atom array for enhancing quantum detection of single photons. We find that the photon reflection by the topological atom array is nonreciprocal, due to broken time-reversal and inversion symmetries. We explicitly show the advantage provided by topology: the realization of large nonreciprocity only requires the weak breaking of time-reversal symmetry. We show that, for the topological atom array with large bandgap, the nonreciprocity is maximal at an atomic spacing $d=3\lambda_0/4$, in which the edge state shows the subradiance with scaling $\propto \exp(-\nu N)$. A topology-enhanced photon absorption from the waveguide takes place for $\Gamma/\Gamma_0\gg 1$, where $\Gamma$ and $\Gamma_0$ are the decays to waveguide and environment, respectively. This topology-protected single-photon detection can be beneficial for quantum computation in superconducting quantum circuits. Our work demonstrates the importance of topology in light-matter interacting phenomena, and sheds new light on topology-protected quantum photonics.

\begin{acknowledgments}The authors thank Prof. J.-M. Raimond for helpful discussions. Y.X.L. is supported by the Key-Area Research and Development Program of GuangDong Province under Grant No. 2018B030326001, the National Basic Research Program (973) of China under Grant No. 2017YFA0304304, and NSFC under Grant No. 11874037. W.N. was supported by the Tsinghua University Postdoctoral Support Program. T.S. acknowledges the support of NSFC No. 11974363. F.N. is supported in part by: NTT Research, Army Research Office (ARO) Grant No. W911NF-18-1-0358, Japan Science and Technology Agency (JST) (via the Q-LEAP program, and the CREST Grant No. JPMJCR1676), Japan Society for the Promotion of Science (JSPS) (via the KAKENHI Grant No. JP20H00134, and the JSPS-RFBR Grant No. JPJSBP120194828), the Asian Office of Aerospace Research and Development (AOARD) (via Grant No. FA2386-20-1-4069), and the Foundational Questions Institute Fund (FQXi) (via Grant No. FQXi-IAF19-06).
\end{acknowledgments}

\end{document}